\documentstyle{article}

\font\tenrm=cmr10
\font\tenit=cmti10
\font\elevenbf=cmbx10 scaled\magstep 1
\font\elevenrm=cmr10 scaled\magstep 1
\font\elevenit=cmti10 scaled\magstep 1

\font\ninerm=cmr9

\newcommand{\newc}{\newcommand}
\newc{\ra}{\rightarrow}
\newc{\lra}{\leftrightarrow}
\newc{\beq}{\begin{equation}}
\newc{\eeq}{\end{equation}}
\newc{\barr}{\begin{eqnarray}}
\newc{\earr}{\end{eqnarray}}

\renewenvironment{thebibliography}[1]
 { \elevenrm
   \begin{list}{\arabic{enumi}.}
    {\usecounter{enumi} \setlength{\parsep}{0pt}
     \setlength{\itemsep}{3pt} \settowidth{\labelwidth}{#1.}
     \sloppy
    }}{\end{list}}

\begin{document}
\begin{center}
\vglue 0.6cm
{
 {\elevenbf        \vglue 10pt
               NEUTRINO PROPERTIES IN A GUT\\
               \vglue 3pt
               ANALYSIS FOR FERMION MASSES\footnote{
\ninerm\baselineskip=11pt Invited talk presented by
J.D. Vergados at the {\it International School of cosmological dark matter
Valencia, SPAIN - 4  Oct. 1993 }
\\}
\vglue 5pt

\vglue 1.0cm
{\tenrm George K. Leontaris and John D. Vergados\\}
\baselineskip=13pt
{\tenit Physics Department, Ioannina University \\}
\baselineskip=12pt
{\tenit Ioannina, GR-451 10, Greece\\}}

\vglue 0.8cm
{\tenrm ABSTRACT}
\end{center}
In practically all extentions of the standard model, the neutrinos
naturally acquire a mass. The neutrino mass matrix, however, contains
many parameters which can neither be predicted by the prevailing
models nor can be fitted to the data.
We propose a  fermion mass matrix ansatz in the context of Grand
Unified Supersymmetric Theories (GUTs) at the GUT scale and use
the renormalization  group equations for the gauge and Yukawa
couplings to predict the 13 arbitrary parameters of
the Standard model. The constraints imposed by the charged
fermion data are
used to  make predictions in the neutrino sector.
The neutrino mass matrix is worked out in the case of the flipped
SU(5) model
 and the predictions are compared with the available experimental data.



\vglue 0.6cm
{\elevenbf\noindent 1. Introduction}
\vglue 0.2cm
\baselineskip=14pt
\elevenrm
Although the Standard Model of Strong and Electroweak interactions
explains all known experimental data, it is now widely accepted that
it is unlikely to be the fundamental theory of nature.

In a fundamental theory the various experimentally measurable parameters
should be
calculable only from few inputs which could be specified in terms
 of the basic principles of the theory.
For example, in String theories\cite{gsw} one expects that all
masses and mixing angles can be determined only from one input, namely the
String gauge coupling $g_{String}$ at the String Unification scale
$M_{String}\sim O(g_{String}\times M_{Pl})$.

In most of the extensions of the standard model the neutrinos naturally
acquire a non-zero mass. The neutrino mass matrix, however, contains
many parameters which can neither be predicted by the preveiling models
nor be fitted to the data. The Grand Unified models (GUTs) in conjunction
with reasonable assumptions  about the higgs representations can lead,
though, to useful relations.
Thus, the isotriplet majorana mass matrix vanishes and the Dirac
neutrino mass matrix usually is connected to the up-quark mass matrix.
The isosinglet majorana  mass matrix for the right handed neutrinos
in most of the GUTs remains unconstrained.

In this talk, I will analyse a particular fermion mass matrix
Ansatz at the GUT scale, which was inspired from the flipped SU(5) model
\cite{barr,aehn}
to predict the charged fermion mass spectrum of the theory at low
energies. We further extend the Ansatz by relating the isosinglet right
handed neutrino mass matrix with that of the down quarks and use
 the constraints imposed by the charged  fermion data
 to  make predictions in the neutrino sector.
The overall scale of the right handed neutrino mass matrix is fitted by
the neutrino oscillation data and the recent COBE data.


\vglue 0.6cm
{\elevenbf\noindent 2. Neutrino Oscillations.}
\vglue 0.2cm

Non-zero masses can lead to many interesting phenomena in physics,
such as lepton flavor and lepton number violation,
as well as the phenomenon of neutrino oscillations.

In most of the GUT models the neutrino masses are predicted to be
very light and usually do not have any significant effect in
flavour changing reactions as $\mu \ra e\gamma $ e.t.c.\cite{KLV} .
In this case the best constraints on
the various suggested models which predict non-zero neutrino masses
may come only from neutrino oscillations.
Neutrino oscillations where suggested more than three decades ago by
Pontecorvo \cite{Po}. They may take place either in vacuum or in the
medium.
\\
{ \it i) neutrino oscillations in vacuum}.\\

Let us for simplicity consider two neutrino flavors.
\beq
\nu_{\alpha} ,\,\,  \nu_{\beta} \,\,   (e.g. \,\,\, \nu_e,  \nu_{\mu})
\label{eq:eg1}
\eeq
These can be written in terms of the mass eigenstates as follows:
\begin{eqnarray}
\nu_{\alpha} = cos\vartheta \nu_1 + sin\vartheta \nu_2 \,\,\,   \\
 \nu_{\beta} = -  sin\vartheta \nu_1 +  cos\vartheta \nu_2
\label{eq:eg2}
\end{eqnarray}
with $\nu_1,\nu_2$ being the  stationaty states.
At a time $t$ one has
\beq
\nu_{\beta}(t) = -  sin\vartheta e^{-iE_1 t} \nu_1 +  cos\vartheta
 e^{-iE_2 t} \nu_2
\eeq
with
\beq
E_i = \sqrt{P^2_{\nu} + m^2_i}  ,  i = 1, 2
\eeq
The probability that $\nu_{\alpha}$ oscillates to $\nu_{\beta}$ is given
by
\beq
P(\nu_{\alpha} \ra  \nu_{\beta}) = |\prec \nu_{\beta}(t)|\nu_{\alpha} (0)
\succ|^2
 = (sin 2\vartheta )^2 sin^2 \Delta_{12}
\eeq
where
\beq
\Delta_{12} = \frac{1}{2}  (\sqrt{P^2_{\nu} + m^2_2} - \sqrt{P^2_{\nu} +
m^2_1}) L, \,\,  L = ct
\eeq

For
$m_i \ll P_{\nu}$,  $i = 1,2$, we have     $ P_{\nu} = E_{\nu}$ and
\beq
\Delta_{12} = \frac{1}{4E_{\nu}} ( m^2_2 -  m^2_1)L = \pi
\frac{L}{\ell_{12}},\,\,\,
 \ell_{12} \equiv  \frac{4 \pi E_{\nu}} {\delta m^2}, \,\,\,
\delta m^2 =  m^2_2 -  m^2_1
\eeq
where
$\ell = 2.476 m\frac{E_{\nu} (MeV)} {\delta m^2(eV^2)}$
 is the oscillation length.
\\
{ \it ii) neutrino oscillations in medium}.\\

Since long time ago, it has been suggested that neutrino oscillations
may offer an explanation to the solar neutrino puzzle. However, the
vacuum oscillations that have been analysed in the previous
subsection do not seem to account for the observed depletion of the
solar electron neutrinos for essencially two reasons:
$ \alpha )$ the expected mixing angle is small, and
$ \beta )$ the oscillation length accidentaly must coinside with the
sun-earth distance.  A more natural explanation may be given
in terms of the so called $MSW$\cite{MSW} effect.  This mechanism
attributes the reduction of the solar neutrino flux which reaches
the earth in the difference between the $\nu_e-e$ and $\nu_{\mu}-e$
(and/or) $\nu_{\tau}-e$ scattering cross sections.

For an arbitrary neutrino state vector let us write
\beq
\Psi (t) = \sum_l |\nu_e\succ \alpha_l(t)
\eeq
Let $\rho_e$ is the electron density of matter. The evolution
equation in flavor space can be written
\beq
\imath \frac{\partial|\alpha \succ}{\partial t}={\cal H}|\alpha \succ
\eeq
with
\begin{eqnarray}
{\cal H}=
&=&\left(\begin{array}{cc}E_{\nu}&sin2\vartheta \frac{\delta m^2}
{4E_{\nu}}\\
sin2\vartheta \frac{\delta m^2}
{4E_{\nu}}&E_{\nu}+cos2\vartheta \frac{\delta m^2}
{2E_{\nu}}
\end{array}
 \right)\nonumber \\
&+&\left(\begin{array}{cc}\alpha_{\nu_e}&0\\
0&a_{\nu_{\alpha}}
\end{array}
 \right) \label{eq:ham}
\end{eqnarray}
where $\alpha_{\nu_e},\alpha_{\nu_{\alpha}}$ are the corresponding
interactions with matter. The quantity of interest here is their
difference which is given by
\beq
\alpha_{\nu_e}-\alpha_{\nu_{\alpha}}\,=\,\sqrt{2} G_F\rho_e(\bf x)
\eeq

The local interaction matrix can be rewritten in a convenient form
as follows
\begin{eqnarray}
 {\cal H}
&=&\left[\begin{array}{cc} E_\nu -\frac{\pi}{\ell} cos2\vartheta +
 \frac{2\pi}{\ell_0(x)} &
\frac{\pi}{\ell} sin 2\vartheta \\
 \frac{\pi}{\ell} sin2\vartheta& E_\nu +\frac{\pi}{\ell}
cos\vartheta
\end{array}\right]
\end{eqnarray}
where $\ell$ and $\vartheta$ were defined above and
$\ell_0(x)=\frac{4\pi}{2\sqrt{2}}
 \frac{1}{G_F \rho_e(\bf {x})}$ is the matter interaction  length.
Thus the local eigenstates for propagation through matter are
\begin{eqnarray}
|\nu_L(x)\succ = cos\vartheta_m(x) |\nu_e\succ -
sin\vartheta_m(x) |\nu_\alpha\succ \,\, (light)\\
 \nu_H(x)\succ = sin\vartheta_m(x) |\nu_e\succ +
cos\vartheta_m(x) |\nu_\alpha\succ \,\, (heavy)
\end{eqnarray}
with
\beq
sin2\vartheta_m(x) = \frac{sin2\vartheta}
{\sqrt{1-2cos2\vartheta \frac{\ell}{\ell_0(x)} +
(\frac{\ell}{\ell_0(x)})^2}}
\eeq
and oscillation length
\beq
\ell_m(x) = \frac{\ell} {\sqrt{1-2cos2\vartheta \frac{\ell}{\ell_0(x)} +
(\frac{\ell}{\ell_0(x)})^2}}
\eeq
Note the presence of resonance at $x_R$ such that
\begin{eqnarray}
H_{ee} = H_{\alpha\alpha}\\
\ell= cos2\vartheta \ell_0(x_R) \,\,(position\,\, of\,\, resonance)
\end{eqnarray}
or
\beq
cos2\vartheta \frac{\delta m^2}{1eV^2} = 0.7 \times 10^{-8}
\frac{\rho(matter)}{gr/cm^3} \frac{E_\nu}{MeV}
\delta m^2
\eeq
thus $\delta m^2 \sim (10^{-9}-10^{-4})eV^2$.
Moreover, at resonance the following conditions hold
\beq
 \ell_m (x_R) = \frac{\ell}{sin2\vartheta}  \gg \ell, \\
   sin2\vartheta_m = 1 \,\,\, (maximal\,\, mixing)
 \eeq

The expression for oscillation probability becomes somewhat complicated.
In the adiabatic approximation becomes
\beq
P(\nu_e(x_0) \ra \nu_e(x)) = \frac {1} {2} [1 + cos2(\vartheta_m(x) -
\vartheta_m(x_0))] - sin 2 \vartheta_m(x) sin 2\vartheta_m(x_0) sin^2\Delta
\eeq
where
\beq
\Delta \simeq \pi \int^x_{x_0} \frac {dx} {\ell_m(x)}
  \eeq
where $x_0$ is the  source point and $ x$ the detection point.
In vacuum $\vartheta_m(x) = \vartheta_m(x_0) = \vartheta$, thus we get
\beq
P(\nu_e \ra \nu_e) = 1 - sin^2 2\vartheta sin^2 \Delta, \,\,\,
\Delta = \pi \frac {L} {\ell}, \,\,\, L = x - x_0
  \eeq

In the case of fast oscillations we get an expression independent of $\Delta$

\beq
P(\nu_e \ra \nu_e) = \frac {1} {2} [1 + cos2(\vartheta_m(x)
cos2(\vartheta_m(x_0)]
\eeq

In the extreme non-adiavatic approximation one finds
\beq
P(\nu_e(x_0) \ra \nu_e(x)) = \frac {1} {2}
[1 + cos2\vartheta_m(x)cos2\vartheta_m(x_0) (1 - 2P_{12})]
  \eeq
where
\beq
P_{12} = e^{- \frac{\pi}{2} \gamma(x_R)}\,\,\,\,  (Landau\,\, Lerner
\,\,Approximation)
  \eeq
with
\beq
\gamma(x_R) = \frac{\delta m^2} {2E_\nu}
\frac{sin^2 2\vartheta_m (x_R)}{cos2(\vartheta_m(x_R)}
\frac {1}{\frac{d\,lnP_l(x)}{d\,x}} \Big|_{x=x_R}
  \eeq


\vglue 0.6cm
{\elevenbf\noindent 3. Structure of Fermion Mass Matrices at  $M_{GUT}$}
 \vglue 0.4cm
The quark and lepton mass matrices can be cast in the following
form at the GUT scale\cite{lv}
\begin{eqnarray}
M_U&=&\left(\begin{array}{ccc}0&0&x\\0&y&z\\x&z&1\end{array}
 \right)\lambda _{top}(t_0) {\upsilon
 \over\sqrt{2}}sin\beta, \label{eq:upq}\\
M_D&=&\left(\begin{array}{ccc}
0&a e^{i\phi}&0\\a e^{-i\phi}&b&0\\0&0&1\end{array} \right)
\lambda _b(t_0){\upsilon
\over\sqrt{2}} cos\beta \label{eq:downq}\\
M_E&=&\left(\begin{array}{ccc}
 0&a e^{i\phi}&0\\a e^{-i\phi}&-3b&0\\0&0&1\end{array} \right)
\lambda _{\tau}(t_0){\upsilon
\over\sqrt{2}} cos\beta \label{eq:elec}
 \end{eqnarray}
with $tan\beta \equiv {<\bar h>
\over <h>}$, $\lambda _b(t_0)=\lambda _{\tau}(t_0)$, and $\upsilon =246 GeV$.
We have taken the up-quark matrix to be symmetric, and the down quark matrix
to be hermitian. Furthermore,  the charged lepton mass matrix is related
in a standard way to the down quark mass matrix at the GUT scale.

 Our ansatz has
a total of five zeros in  the quark sector (the leptonic sector is directly
related to them);  two zeros for the up and three for the down quark mass
 matrix(zeros in symmetric entries are counted only once). These zeros reduce
the number of arbitrary parameters at the GUT-scale to eight, namely
$x,y,z,a,b,\phi,\lambda _b(t_0)$, and $\lambda _{top}(t_0)$. These
non-zero entries should serve to determine the thirteen arbitary parameters
of the standard model, i.e., nine quark and lepton masses, three mixing
angles and the phase of the Cabbibo-Kobayashi-Maskawa (CKM) matrix.
Thus, as far as the charged fermion sector is concerned, we end up with five
predictions (we have previously discussed the predictions for the neutrino
sector
in \cite{lv}). We may reduce the number of arbitary parameters
by one, if we impose more structure in the up-quark mass matrix.
We may for example relate the $(13)$, $(23)$ and $(22)$ entries, as follows
\begin{eqnarray}
y =  n z^2 {},{} x =  (n-1) z^2
\end{eqnarray}
where $n$ as will be shown, is a number in the range $n\sim (3-6)$.
Although this structure is imposed by hand \cite{lv},
it increases the predictive power of the theory,
as it reduces the number of free parameters by one.

  In order to find the  structure of the mass matrices at
the low energy scale and calculate the mass eigenstates as well as
the mixing matrices and compare them  with the experimental data, we
need to evolve them down to $m_W$, using the renormalization group
equations. The RGEs for the Yukawa couplings at the one-loop level, read
\begin{eqnarray}
16\pi^2 \frac{d}{dt} \lambda_U&=& \left( I\cdot
Tr [3 \lambda_U\lambda_U^\dagger ]  +
3 \lambda_U \lambda_U^\dagger +\lambda_D \lambda_D^\dagger
-I\cdot G_U\right) \lambda_U, \label{eq:rge1}
\\
16\pi^2 \frac{d}{dt} \lambda_D&=& \left( I\cdot
Tr [3 \lambda_D\lambda_D^\dagger +
\lambda_E \lambda_E^\dagger ]  +
3 \lambda_D \lambda_D^\dagger +\lambda_U \lambda_U^\dagger
-I \cdot  G_D\right) \lambda_D, \label{eq:rge2}
\\
16\pi^2 \frac{d}{dt} \lambda_E&=& \left( I\cdot
 Tr [ \lambda_E\lambda_E^\dagger +3
\lambda_D  \lambda_D^\dagger ]  + 3
 \lambda_E \lambda_E^\dagger -I \cdot G_E\right) \lambda_E,
\label{eq:rge3}
\end{eqnarray}
where $\lambda_\alpha$, $\alpha=U,D,E$, represent the $3$x$3$ Yukawa matrices
which are defined in terms of the mass matrices given in
Eqs(\ref{eq:upq}-\ref{eq:elec}), and $I$ is the $3$x$3$ identity matrix.
$t\equiv\ln(\mu/\mu_0)$, $\mu$ is the scale at which
the couplings are to be
determined and $\mu_0$ is the reference scale,
in our case the GUT scale. The
gauge contributions are given by
\begin{eqnarray}
G_\alpha&=&\sum_{i=1}^3 c_\alpha^i g_i^2(t),\\
g_i^2(t)&=&\frac{g_i^2(t_0)}{1- \frac{b_i}{8\pi^2} g_i^2(t_0)(t-t_0)}.
\end{eqnarray}
The $g_i$ are the three gauge coupling constants of the Standard Model and
$b_i$
are the corresponding supersymmetric beta functions. The coefficients
$c_\alpha^i$ are given by
\begin{eqnarray}
\{c_Q^i \}_{i=1,2,3} = \left\{ \frac{13}{15},3,\frac{16}{3} \right\}, \qquad
\{c_D^i \}_{i=1,2,3} = \left\{\frac{7}{15},3,\frac{16}{3} \right\},\\
\{c_U^i \}_{i=1,2,3} = \left\{ \frac{16}{15},0,\frac{16}{3}\right\}\qquad
\{c_E^i \}_{i=1,2,3} = \left\{ \frac{9}{5},3,0\right\} .
\end{eqnarray}
In what follows, we will assume that $\mu_0\equiv M_G\approx 10^{16}GeV$
and $a_i(t_0)\equiv a_G \equiv \frac{g_i^2(t_0)}{4\pi}\approx
\frac{1}{25}$. In order to evolve the Equations
(\ref{eq:rge1}-\ref{eq:rge3}) down to low energies, we also need to do
some plausible approximations. For later convenience, we define a new
parameter $tan\theta_1=z$ and
 diagonalize the up quark mass matrix at the GUT scale and obtain
the eigenstates \begin{eqnarray}
m_1(M_G)\approx -n(n-1)ptan^2\theta_1 sin^2\theta_1 \nonumber\\
m_2(M_G)\approx (n-1)ptan^2\theta_1\label{eq:Gmass}\\
m_3(M_G)\approx \frac{p}{cos^2\theta_1}\nonumber
\end{eqnarray}
with diagonalizing matrix
\begin{eqnarray}
{\it K}&=&\left(\begin{array}{ccc}
\frac{1}{{\it D_1}}&\frac{-sin\theta_1}{{\it D_2}}&\frac{(n-1)sin^2\theta_1}
{{\it D_3}}\\
\frac{sin2\theta_1}{2{\it D_1}}&\frac{1}{{\it D_2}}&\frac{sin\theta_1}
{{\it D_3}}\\
\frac{nsin^2\theta_1}{{\it D_1}}&\frac{sin\theta_1}{{\it D_2}}&
\frac{1-nsin^2\theta_1}{{\it D_3}}
\end{array} \right) \label{eq:upk}
\end{eqnarray}
where we have set $p=\lambda_t(t_0){\upsilon \over\sqrt{2}}sin\beta $ and ${\it
D_1}\approx \sqrt{ 1+sin^2\theta_1 cos^2\theta_1}$, ${\it D_2}\approx
\sqrt{1+sin^2\theta_1 }$, and  ${\it D_3}\approx \sqrt{1-(2n-1)sin^2\theta_1
}$.
In this case all other mass matrices are also rotated by the same similarity
transformation, thus
\begin{eqnarray}
\lambda_{\alpha}&\rightarrow &{\tilde\lambda}_{\alpha}=
 K^\dagger\lambda_{\alpha}K,\alpha =D,E
\end{eqnarray}
 Now, assuming that the $\lambda_{top}$ coupling is much bigger
than all other fermion Yukawa couplings, we may ignore the contributions of
the latter in the RHS of the RGEs in Eqs(\ref{eq:rge1}-\ref{eq:rge3}).
In this case all differential equations reduce to a simple uncoupled form.
Thus the top-Yukawa coupling differential equation, for example,
may be cast in the form
\begin{eqnarray}
16\pi^2 \frac{d}{dt}
\tilde\lambda_{top}&=&\tilde\lambda_{top}(6{\tilde\lambda}_{top}
^2-G_Q(t))\label{eq:topeq} \end{eqnarray}
with the solution
\begin{eqnarray}
\tilde \lambda_{top}&=&\tilde \lambda_{top}(t_0)\xi^6\gamma_Q(t)\
\label{eq:ltop}
\end{eqnarray}
where
\begin{eqnarray}
\gamma_Q(t)&=& \prod_{j=1}^3 \left(1- \frac{b_{j,0}\alpha_{j,0}(t-t_0)}{2\pi}
\right)^{c_Q^j/2b_j},\\
\xi&=& \left( 1-\frac{6}{8\pi^2}\lambda_{top}^2(t_0)
\int_{t_0}^{t} \gamma_Q^2(t)\,dt \right)^{-1/12}\label{eq:ksi}.
\end{eqnarray}
Thus the GUT up quark mass eigenstates given in Eq(\ref{eq:Gmass})
 renormalized down to their mass scale, are given by
\begin{eqnarray}
m_u \approx \gamma_Q\xi^3 \eta_u n(n-1)ptan^2\theta_1 sin^2\theta_1\nonumber\\
m_c \approx \gamma_Q\xi^3\eta_c (n-1)ptan^2\theta_1\label{eq: upmass}\\
m_t \approx \gamma_Q\xi^6 \frac{p}{cos^2\theta_1}\nonumber
\end{eqnarray}
where $\eta_u$ , $\eta_c$ are taking into account the renormalization
effects  from the $m_t$-scale down to the mass of the corresponding quark.
In the following we will take $\eta_u(1GeV) = \eta_c(1GeV)\approx 2$.
We may combine  Eqs(\ref{eq: upmass}) and give predictions
for the top quark mass and  the mixing angle $\theta_1 $
in terms of the low energy up and charm quark masses. We obtain
\begin{eqnarray}
m_t \approx
\xi^3\frac{n}{n-1}\frac{m_c^2}{m_u}\frac{\eta_u}{\eta_c^2}\label{eq: tpred}\\
sin\theta_1 =\sqrt{\frac{m_u}{nm_c}}\label{eq:thpred}
\end{eqnarray}
In the basis where the up-quark mass matrix is diagonal,
the renormalized down-quark and charged lepton mass matrices are given by
\begin{eqnarray}
m_D^{ren} &\approx & \gamma_D I_{\xi}{\it K^{\dagger}}
m_D{\it K}\label{eq:rdown}\\
m_E^{ren} &\approx & \gamma_E {\it K^{\dagger}}m_E{\it K} \label{eq:relec}
\end{eqnarray}
where $I_{\xi}=Diagonal(1,1,\xi)$,
and $m_{D,E}$, are given in Eqs(\ref{eq:downq}) and (\ref{eq:elec}).
In order to make definite predictions in the fermion mass sector, we choose
the lepton masses as inputs and express the arbitrary parameters
$a,b,\lambda_{\tau}(\equiv \lambda_b) $ in $m_E$ in terms of $m_e,m_{\mu}$,
and $m_{\tau}$. Substituting into $m_D$ and diagonalizing $m_D^{ren}$, we
obtain\cite{lv}
\begin{eqnarray}
m_d\approx  6.3\times (\frac{\eta_d}{2})MeV \\
m_s\approx 153 \times (\frac{\eta_s}{2})MeV \\
m_b \approx \eta_b \frac{\gamma_D}{\gamma_E}\xi m_{\tau}\label{eq:bottom}
\end{eqnarray}
Since $\eta_d\approx \eta_s\approx 2$, the predictions for the light quarks
$m_{d,s}$ are within the expected range\cite{qcd}.
Now, in order to make a prediction for the bottom quark, we need to know
the value of $\xi$, but the latter depends on the top quark
coupling at the GUT
scale  as well as on the top-mass, through Eq.(\ref{eq:ksi}).
 We can use, however,
Eq.(\ref{eq:bottom}), to predict the range of the top mass.
Thus, using the available limits
on the bottom mass, $4.15 \le m_b\le 4.35 GeV$, and $\eta_b\approx 1.4$,
 we can obtain the
following range for $m_{top}$
\begin{eqnarray}
125GeV\le m_{top}  \le 170 GeV\label{eq:toprange}
\end{eqnarray}

As we work in a diagonal basis for the up-quark mass matrix,
the CKM - matrix
can be found by diagonalizing $m_D^{ren}$ in Eq(\ref{eq:rdown}).
Let us denote with $U_{\theta_2}$ the matrix which diagonalizes the
$m_D(M_{G})$: \begin{eqnarray}
U_{\theta_2}&=&\left(\begin{array}{ccc}
cos\theta_2e^{-\imath\phi}&-sin\theta_2&0\\sin\theta_2&cos\theta_2e^
{-\imath\phi}&0 \\0&0&1\end{array} \right) \label{eq:ckm}
\end{eqnarray}
Then the angle $\theta_2$,
 can be determined in terms of the lepton masses making
use of the relations between $m_D$ and $m_E$ entries at the GUT scale. We find,
$sin\theta_2\approx \frac{m_e}{m_{\mu}} $. The CKM-matrix can now be expressed
to a good approximation by  $V_{CKM}\approx
U_{\theta_2}I_{\xi}K^{\dagger}I_{\xi}^{-1}U_{\theta_2}^{\dagger}$. In
particular,
 we get the following predictions for $V_{cd}$,$V_{ts}$ and $V_{td}$:
 \begin{eqnarray}
V_{cd}\approx  (-s_1c_2e^{\imath\phi}-s_2)/D_1\\
V_{ts}\approx  \xi s_1((n-1)s_1c_2+e^{\imath\phi}c_2)/D_3\\
V_{td}\approx  \xi s_1((n-1)s_1c_2-s_2)/D_3\label{eq:KMij}
\end{eqnarray}
where $c_i=cos\theta_i$ and $s_i=sin\theta_i$,$i=1,2$. As we have shown, both
angles are given in terms of well known low energy fermion masses, while from
the experimental range of $V_{ij}$ the parameter $n$ is constrained to be in
the range $3\le n \le 6$.

 As an example, we give the CKM-entries for
the specific
case where, $m_t=135GeV$, $tan\beta =1.1$,and $\phi =\frac{\pi}{4}$.
In this case $m_b$ is
predicted to be $4.33GeV$, while we obtain
\begin{eqnarray}
{\mid V_{CKM} \mid}_{ij}&=&\left(\begin{array}{ccc}
0.9754&0.22&0.0032\\0.22&0.9750&0.038\\0.01&0.037&0.9993\end{array} \right)
 \label{eq:ckm}
\end{eqnarray}


\vglue 0.6cm
{\elevenbf\noindent 4. Neutrino Masses}
 \vglue 0.4cm
There is a lot of experimental evidence today, that the neutrinos have non-zero
masses.
For example, recent data from solar neutrino experiments\cite{H1}
 show that the
deficiency of solar neutrino flux, i.e. the discrepancy
between theoretical estimates and the experiment, is
naturally explained if the $\nu _e$ neutrino oscillates to
another species during its flight to the earth.
Furthermore, the COBE
measurement \cite{Smooth} of the large scale
microwave background anisotropy, might be
explained \cite{SShafi} if one assumes an
admixture of cold ($ \sim 75\% $) plus hot ($\sim25\% $)
dark-matter. It is hopefully expected that some
 neutrino (most likely $\nu _{\tau }$)
  may be the natural candidate of the hot dark
matter component.

Here we would like to address the question of
neutrino masses in GUT models
which arise \cite{aehn,al}in the free fermionic construction of
four dimensional strings.  Taking into
account renormalization effects ,
 it was recently shown\cite{eln,lv} that the
general see-saw mechanism which occurs naturally in the flipped
model, turns out to be consistent with the recent solar neutrino
data, while on the other hand suggests that CHOROUS and NOMAD
experiments at CERN may have a good chance of observing
$\nu_{\mu}\longleftrightarrow  \nu_{\tau }$ oscillations\footnote{
\ninerm\baselineskip=11pt for neutrinos in convensional GUTs
 see\cite{various}}

The various tree-level superpotential mass terms which contribute to
the neutrino mass matrix in the flipped $SU(5)$ model are the
following:
\begin{eqnarray}
\lambda _{ij}^{u}F^{i}\bar f^{j}\bar h +\lambda
_{ij}^{\phi \nu ^{c}}F^{i}\bar H\phi ^{j}+ \lambda _{ij}^{\phi}\phi ^0\phi
^{i}\phi ^{j} \label{eq:potential}
\end{eqnarray}
where in the above terms $F^{i},\bar f^{j}$ are the $ {10},\bar 5$
matter SU(5) fields while $\bar H,\bar h, h$ are the $\bar {10},\bar
5, 5$ Higgs representations and $\phi^{i}$ are neutral $SU(5)\times
U(1)$  singlets. The Higgs field $\bar H$ gets a vacuum expectation
value(v.e.v.) of the order of the SU(5) breaking scale ($\sim
10^{16}GeV$), $\bar h, h$ contain the standard higgs doublets while
$\phi ^0$ acquires a v.e.v., most preferably at the electoweak scale.
The neutrino mass matrix may also receive significant contributions
from other sources. Of crusial importance, are the
non-renormalizable contributions which may give a direct isosinglet
$M_{\nu ^c\nu ^c}=M^{rad}$ contribution which is absent in the
tree-level potential. Then, the general $9\times 9$ neutrino mass
matrix in the basis $(\nu _i,\nu _i^c,\phi _i)$, may be written as
follows:

\begin{eqnarray}
m_{\nu }=
\left(\matrix{0&m_{\nu_D}&0\cr
m_{\nu_D}^{\dagger}&M^{rad}&M_{\nu}^{c,\phi }\cr
0&M_{\nu }^{c,\phi\dagger} &\mu _{\phi }\cr}\right) \label{eq:SeeSaw}
\end{eqnarray}
where it is understood that all entries in Eq.(\ref{eq:SeeSaw}) represent
 $3\times 3$ matrices. The above neutrino matrix is different from that of
standard see-saw matrix-form, since now there are three neutral
$SU(5)\times U(1)$ singlets involved, one for each family.

It is clear that the matrix (\ref{eq:SeeSaw}) depends on a relatively
 large number of
parameters and a reliable estimate of the light neutrino masses and the
mixing angles  is a rather complicated task. We are going to use however
 our knowledge of the rest of the fermion spectrum to reduce
sufficiently the number of parameters involved. Firstly, due to the GUT
relation $m_U(M_{GUT})=m_{\nu _D}(M_{GUT})$, we can deduce the form of
$m_{\nu _D}(M_{GUT})$, at the GUT scale in terms of the up-quark masses.
The heavy majorana $3\times 3$ matrix $M^{rad}$, depends on the kind of the
specific generating mechanism. Here\cite{lv} we take it to be proportional to
th
e
down quark-matrix at the GUT scale:
\begin{eqnarray}
M^{rad}=\Lambda ^{rad} m_D(M_{GUT})\label{eq:mrad}
\end{eqnarray}
 The
$M_{\nu ^{c},\phi }$ and $\mu _{\phi }$  $3\times 3$ submatrices are
also model dependent. In most of the string models however, there is
only one entry at the trilinear superpotential in the matrix $M_{\nu
^{c},\phi }$, which is of the order $M_{GUT}$. Other terms, if any,
usually arise from high order non-renormalizable terms. We will
assume in this work  only the existence of the trilinear term,
since higher order ones will be comparable to  $M^{rad}$  and are
not going to change our predictions. In particular we will take
$M_{\nu ^{c},\phi }\sim Diagonal(M,0,0)$, and $\mu _{\phi }\sim
Diagonal(\mu,\mu^{\prime},\mu^{\prime \prime})$,
with $\mu ,...<< M\sim M_{GUT}$, thus we will treat
 Eq.(\ref{eq:SeeSaw}) as a $7\times 7$ matrix.

To obtain the neutrino spectrum and lepton mixing, we must
introduce values for the two additional papameters $M, \Lambda
^{rad} $ of the neutrino mass matrix in Eq.(\ref{eq:SeeSaw}).
We assume naturally
$M=<\bar H>\approx 10^{16}GeV$.
The neutrino mass eigenvalues can now be predicted in terms of the scale
 quantity
$\Lambda ^{rad}$ . Thus they can be written as\cite{lv}
\begin{eqnarray}
m_{\nu _e}\approx 0,
m_{\nu _{\mu}}={\Lambda ^{\mu}\over \Lambda
^{rad}}\times 10^{-2}eV,
m_{\nu _{\tau}}={\Lambda ^{\tau}\over \Lambda
^{rad}}\times 10eV\label{eq:eigeneut}
\end{eqnarray}
For $m_t\approx 130GeV$ we get $\Lambda ^{\mu} \approx .80\times
10^{12}$ and  $\Lambda ^{\tau} \approx 1.85\times 10^{12}$.

For the oscillation probabilities, we find\cite{lv}
\begin{eqnarray}
P(\nu _e\leftrightarrow \nu _\mu ) \approx 3.1\times 10^{-2}
sin^2({\pi {L\over l_{12}}})\label{eq:osc12}
\end{eqnarray}

 \begin{eqnarray}
 P(\nu _{\tau}\rightarrow \nu _{\mu }) \approx
4.0\times 10^{-3} sin^2({\pi {L\over l_{13}}})\label{eq:osc23}
\end{eqnarray}
\begin{eqnarray}
 P(\nu _e\rightarrow \nu _{\tau }) \approx
4.0\times 10^{-5} sin^2({\pi {L\over l_{13}}})\label{eq:osc13}
\end{eqnarray}
where L is the source--detector distance and
\begin{eqnarray}
l_{ij}={4\pi E_{\nu}\over |m_i^2-m_j^2|}\label{eq:oscl}
\end{eqnarray}

We can determine the range of
$\Lambda ^{rad}$, using the available solar neutrino data
(see for example ref\cite{petcov} for a systematic analysis of the allowed
regions using all available data). From such data, we extract for the
region of small mixing angle
 \begin{eqnarray}
5.0\times 10^{-3}\le sin^22\theta_{ij}\le 1.6\times
10^{-2}\label{eq:data1}
\end{eqnarray}
\begin{eqnarray}
0.32\times 10^{-5}(eV)^2\le \delta m_{ij}^2\le 1.2\times
10^{-5}(eV)^2\label{eq:data2}
\end{eqnarray}
 Our result in Eq(\ref{eq:osc12})
is a bit outside the above range but the mass constraint can be
easily satisfied by choosing $\Lambda ^{rad}$ in the range
\begin{eqnarray}
.7\times 10^{12}\le \Lambda ^{rad}\le 7\times 10^{12}\label{eq:lamrad}
\end{eqnarray}
 Our neutrino masses can also  easily be made to fall into the range
of the Frejus atmospheric neutrinos
\begin{eqnarray}
10^{-3}(eV)^2\le \delta m_{ij}^2\le 10^{-2}(eV)^2
\end{eqnarray}
but our mixing is much too small. Our results are also consistent
with the data on $\nu _{\mu } \leftrightarrow \nu _{\tau } $
 oscillations\cite{mutau}
\begin{eqnarray}
 sin^22\theta_{\mu \tau}\le 4.\times 10^{-3},\delta m_{\nu _{\mu}
\nu _{\tau}}^2\ge 50(eV)^2\label{eq:mtosc}
\end{eqnarray}
Our results however cannot be made to fall on the $sin^22\theta $
$vs$  $\delta m^2$ of the $BNL$ $\nu _{\mu } \leftrightarrow \nu _e $
oscillation results\cite{boro}, so long as we insist in
Eq.(\ref{eq:mrad})

Moreover, with the quantity $\Lambda^{rad}$ given by Eq.(\ref{eq:lamrad}),
we obtain $m_{\nu _{\tau }}\approx
(2 \sim 20)eV$, hence one can obtain simultaneously the
cosmological HOT-dark matter component,
 in agreement with the interpretation of the COBE data.
 Indeed
an upper limit on the $\nu _{\tau }$ mass can be obtained from
the formula
\begin{eqnarray}
7.5\times 10^{-2}\le \Omega _{\nu }h^2\le 0.3\label{eq:ntau1}
\end{eqnarray}
Translating this into a constraint on $m_{\nu {_\tau}}$, arising
from the relation $m_{\nu_{\tau }}\approx \Omega _{\nu }h^291.5eV$
where $h=.5\sim 1.0$ is the Hubble parameter, one gets the range
\begin{eqnarray}
6.8\le m_{\nu _{\tau }}\le 27eV\label{eq:ntau1}
\end{eqnarray}
 which can be easily achieved with the above range of $\Lambda
^{rad}$.
 \vglue 0.2cm
{\elevenbf Acknowledgements} {\elevenit J.D.V. would like to thank
the organizing committee of the International School on Cosmological
Dark Matter
 for their kind hospitality}.
\vglue 0.3cm

\newpage
 {\elevenbf\noindent 5. References \hfil}  \vglue 0.4cm

\end{document}